\documentclass[twocolumn,showpacs,preprintnumbers,amsmath,amssymb]{revtex4}

\usepackage{graphicx}
\usepackage{dcolumn}
\usepackage{bm}

\begin{document}

\title{Quantum Cosmology for Tunneling Universes}

\author{Sang Pyo Kim}\email{sangkim@kunsan.ac.kr}

\affiliation{Department of Physics, Kunsan National University,
Kunsan 573-701, Korea}

\date{\today}
\begin{abstract}
In a quantum cosmological model consisting of a Euclidean region
and a Lorentzian region, Hartle-Hawking's no-bounary wave
function, and Linde's wave function and Vilenkin's tunneling wave
function are briefly described and compared with each other. We
put a particular emphasis on semiclassical gravity from quantum
cosmology and compare it with the conventional quantum field
theory in curved spacetimes. Finally, we discuss the recent debate
on catastrophic particle production in the tunneling universe
between Rubakov and Vilenkin within the semiclassical gravity.
\end{abstract}

\maketitle

\section{Introduction}

It is widely known that quantum gravity effects are important
at the Planck scale of length, time and mass
\begin{eqnarray}
l_p &=& \Bigl(\frac{G \hbar}{c^3} \Bigr)^{1/2} = 10^{-33} ~{\rm cm}, \nonumber\\
t_p &=& \Bigl(\frac{G \hbar}{c^5} \Bigr)^{1/2} = 10^{-43} ~{\rm sec},
\nonumber\\
m_p &=& \Bigl(\frac{c \hbar}{G} \Bigr)^{1/2} = 10^{19} ~{\rm Gev}/c^2.
\end{eqnarray}
It is also well accepted that at such Planck scales quantum
gravity theory should replace the role of classical gravity
theory. Though at present any viable theory of quantum gravity is
not known free from all conceptual and technical problems, quantum
gravity may be denoted symbolically as
\begin{equation}
 \hat{G}_{\mu \nu} = \frac{8 \pi}{ m_p^2} \hat{T}_{\mu \nu}.
\label{qg1}
\end{equation}
It is within quantum gravity with both gravity and matter
quantized that the unitarity problem and renormalization problem,
quantum black holes, and quantum cosmology should be properly
addressed.

In a regime where gravity becomes classical but matter still
maintains quantum nature, one has semiclassical gravity theory
\begin{eqnarray}
G_{\mu \nu} = \frac{8 \pi}{ m_p^2} \langle \hat{T}_{\mu \nu}
\rangle. \label{sqg2}
\end{eqnarray}
Gravity may undergo, for instance, a decoherence mechanism to make
a quantum-to-classical transition. Quantum field theory in curved
spacetimes or the semiclassical theory derived from quantum
gravity may be regarded as semiclassical gravity theory. This
semiclassical gravity theory has been widely used to investigate
important phenomena such as particle and/or entropy production,
black hole information, and some of cosmological problems.
Finally, when both gravity and matters become classical, they are
governed by classical gravity theory
\begin{equation}
G_{\mu \nu} = \frac{8 \pi}{ m_p^2} T_{\mu \nu}.
\label{cg3}
\end{equation}

Even with a consistent theory of quantum gravity known, one should
also understand the mechanism leading quantum gravity (\ref{qg1})
to semiclassical gravity (\ref{sqg2}) and finally to classical
gravity (\ref{cg3}). The main aim of this talk is to follow this
scheme, discuss and clarify some issues in this direction
\cite{sem grav,kim-cqg}. As we do not know the consistent quantum
gravity, we need to adopt some quantum gravity theory that
incorporates quantum gravity effects. The Wheeler-DeWitt (WDW)
equation, for instance, may be such a quantum gravity theory where
both gravity variables from metric tensors and matters are
quantized. Quantum cosmology, where cosmological models are
quantized \`{a} la the WDW equation, can be a good arena to test
the above scheme of a transition of quantum gravity down to
classical gravity.

Quantum cosmology consists of a quantum dynamical law, boundary
(initial) conditions and prediction (interpretation) of quantum
states \cite{page}. The WDW equation as a quantum dynamical law
determines the wave functions of the universe. There should be
some prescription to select a unique wave function of the
universe. A boundary condition selects the unique or appropriate
state of the universe. Then the selected wave function carries all
information of the history of the universe. Three leading
proposals to prescribe boundary conditions were advanced:
Hartle-Hawking's no-boundary wave function \cite{hartle-hawking},
Linde's wave function \cite{linde}, and Vilenkin's tunneling wave
function \cite{vilenkin}. We derive the semiclassical gravity from
the WDW equation and discuss particle creation in a tunneling
universe. Recently there has been a debate on catastrophic
particle creation from a tunneling universe between Rubakov's
group \cite{rubakov} and Vilenkin's group \cite{vilenkin2}. We
show that an adiabatic vacuum state minimizes particle creation
but its squeezed states necessarily lead to catastrophic particle
creation.

\section{Wave functions of the Universe}

As a simple model for quantum cosmology, we consider the
Friedmann-Robertson-Walker (FRW) universe with the action
\begin{eqnarray}
I_g = \frac{m_p^2}{16 \pi} \int d^4 x \sqrt{-g} \Bigl[R - 2
\Lambda \Bigr]
+  \frac{m_p^2}{8 \pi} \int d^3 x \sqrt{h} K,
\end{eqnarray}
where $\Lambda$ is a cosmological constant. The surface term is
introduced to yield the correct equation of motion for a closed
universe. In the Arnowitt-Deser-Misner (ADM) formulation with the
metric
\begin{equation}
ds^2 = -N^2 (t) dt^2 + a^2 (t) d \Omega_3^2,
\end{equation}
the action leads to the Hamiltonian constraint
\begin{equation}
H_g = - \frac{1}{3 \pi m_p^2 a} \pi_a^2 - \frac{3 \pi m_p^2}{4}
\Bigl(a - \frac{\Lambda}{3} a^3 \Bigr) = 0,
\end{equation}
where the conjugate momentum is given by
\begin{equation}
\pi_a = - \frac{3 \pi m_p^2 a}{2 N} \frac{\partial a}{\partial t}.
\end{equation}
The quantum theory of the universe is given by the Hamiltonian
constraint \`{a} la the Dirac quantization, which is nothing but
the WDW equation
\begin{eqnarray}
\Big[- \frac{\hbar^2}{2 m_p^2} \frac{d^2}{d a^2} + V_g (a) \Bigr]
\Psi(a) = 0,  \label{wdw eq}
\end{eqnarray}
with an effective gravitational potential
\begin{eqnarray}
 V_g (a) = \frac{9 \pi^2 m_p^2}{8}
\Bigl(a - \frac{\Lambda}{3} a^4 \Bigr). \label{grav pot}
\end{eqnarray}

The WDW equation for pure gravity of the FRW universe now has the
form of one-dimensional Schr\"{o}dinger equation with the
potential (\ref{grav pot}) and zero energy. No exact solutions are
known for this apparently simple equation. More significantly, the
boundary conditions for quantum cosmology may differ from those
for quantum mechanics. However, the characteristic behaviors of
the wave functions of leading proposals can be well exhibited by
the asymptotic solutions \cite{kim-cos}. For that purpose, we
introduce the Liouville-Green transformation \cite{kim-page}
\begin{equation}
\eta^{3/2} = \frac{3}{2} \int_{a_t}^a da \Bigl( - \frac{2
m_p^2}{\hbar^2} V_g (a) \Bigr)^{1/2}, \label{eta}
\end{equation}
and write the gravitational wave function in the form
\begin{equation}
\Bigl[\frac{d^2}{d \eta^2} + \eta + \Delta
(\eta)  \Bigr] \Psi (\eta) = 0, \label{grav eq}
\end{equation}
where
\begin{equation}
\Delta (\eta) = - \frac{3}{4} \frac{\bigl(d^2 \eta/d a^2
\bigr)^2}{\big(d \eta/d a \bigr)^4} +
\frac{1}{2} \frac{\bigl(d^3 \eta/d a^3
\bigr)}{\big(d \eta/d a \bigr)^3}.
\end{equation}
Here $a_t$ is the turning point, $a < a_t = (\Lambda/3)^{1/2}$,
from the side of a de Sitter region. The Liouville-Green
transformation has an advantage that it can readily be extended to
classically forbidden regions without singular behavior at turning
points. The integral (\ref{eta}) can be done exactly to yield
\begin{eqnarray}
\eta &=& \Bigl( \frac{9 \pi m_p^2}{4 \hbar \Lambda} \Bigr)^{2/3}
\Bigl(\frac{\Lambda}{3} a^2 - 1 \Bigr), \nonumber\\
\Delta (\eta) &=& - \frac{1}{16} \frac{1}{\Bigl[ \eta +
\Bigl(\frac{9 \pi m_p^2}{4 \hbar \Lambda} \Bigr)^{2/3} \Bigr]^2}.
\end{eqnarray}

We can find asymptotic solutions to Eq. (\ref{wdw eq}) at least in
two asymptotic regions, a de Sitter region and a classically
forbidden region. In the first case of the de Sitter region,
$(\eta \gg 1, \Delta = -/ 16 \eta^2 \rightarrow 0)$, Eq.
(\ref{grav eq}) has the asymptotic solutions
\begin{eqnarray}
\Psi (a) = \begin{cases} Ai \bigl(- \eta - \delta (\eta) \bigr) \cr
Bi \bigl( - \eta - \delta (\eta) \bigr) \end{cases}, \label{asym1}
\end{eqnarray}
where $Ai$ and $Bi$ are Airy functions and $\delta$ is a small
correction due to $\Delta$. In the second case of the classically
forbidden region of tunneling universe, we analytically continue
Eq. (\ref{eta}) to introduce another variable
\begin{equation}
\zeta^{3/2} = \frac{3}{2} \int_{a}^{a_t} da \Bigl( \frac{2 m_p^2}{\hbar^2}
V_g (a) \Bigr)^{1/2}, \label{zeta}
\end{equation}
and write the gravitational field equation as
\begin{equation}
\Bigl[\frac{d^2}{d \zeta^2} - \zeta + \Delta
(\zeta)  \Bigr] \Psi (\zeta) = 0, \label{euc grav eq}
\end{equation}
where
\begin{equation}
\Delta (\zeta) = - \frac{3}{4} \frac{\bigl(d^2 \zeta/d a^2
\bigr)^2}{\big(d \zeta/d a \bigr)^4} +
\frac{1}{2} \frac{\bigl(d^3 \zeta/d a^3
\bigr)}{\big(d \zeta/d a \bigr)^3}.
\end{equation}
Now Eq. (\ref{zeta}) yields
\begin{eqnarray}
\zeta &=& \Bigl(\frac{9 \pi m_p^2}{4 \hbar \Lambda} \Bigr)^{2/3}
\Bigl(1 - \frac{\Lambda}{3} a^2 \Bigr), \nonumber\\
\Delta (\zeta) &=& - \frac{1}{16} \frac{1}{\Bigl[ \zeta +
\Bigl(\frac{9 \pi m_p^2}{4 \hbar \Lambda} \Bigr)^{2/3} \Bigr]^{2}}.
\end{eqnarray}
The asymptotic solutions to Eq. (\ref{euc grav eq}) in the
tunneling region are
\begin{eqnarray}
\Psi (a, \phi) =  \begin{cases} Ai \bigl( \zeta - \delta (\zeta)
\bigr) \cr
Bi \bigl( \zeta - \delta (\zeta) \bigr) \end{cases}. \label{asym2}
\end{eqnarray}

Using the asymptotic solutions (\ref{asym1}) and (\ref{asym2}), we
can prescribe the three leading proposals for the wave functions
of the universe. First, Hartle-Hawking's no-boundary wave function
in the classically forbidden region is given by
\begin{equation}
\Psi_{HH} (a) = C Ai \bigl( \zeta - \delta(\zeta) \bigr),\label{hh1}
\end{equation}
and has the asymptotic form
\begin{equation}
\Psi_{HH} (a) = C \Bigl(\frac{1}{1 - \frac{\Lambda}{3} a^2} \Bigr)^{1/4}
e^{- \frac{1}{\hbar} S_g (a)}.
\end{equation}
Here $S_g$ is the gravitational instanton
\begin{equation}
S_g (a) =  \int_a^{a_t} \bigl(2 m_p^2 V_g (a) \bigr)^{1/2} =
\frac{2 \hbar}{3} \zeta^{3/2}.
\label{grav ins}
\end{equation}
In the de Sitter region we analytically continue $\zeta$ to $\eta$
to obtain
\begin{equation}
\Psi_{HH} (a) = C Ai \bigl( - \eta - \delta(\eta) \bigr). \label{hh2}
\end{equation}
In fact, the wave function (\ref{hh2}) is an analytical continuation of
(\ref{hh1}) through the turning point without having any singularity.
The Hartle-Hawking wave function has the asymptotic form
\begin{equation}
\Psi_{HH} (a) = \Bigl(\frac{1}{\frac{\Lambda}{3} a^2 - 1} \Bigr)^{1/4}
\sin \Bigl(\frac{1}{\hbar} S_g (a) + \frac{\pi}{4}\Bigr),
\end{equation}
where
\begin{equation}
S_g (a) = \int_{a_t}^a \bigl(- 2 m_p^2 V_g (a) \bigr)^{1/2}
= \frac{2 \hbar}{3} \eta^{3/2}.
\end{equation}

Second, Linde's wave function is prescribed by
\begin{equation}
\Psi_{L} (a) = C Bi \bigl( \zeta - \delta(\zeta) \bigr),
\end{equation}
and has the asymptotic form
\begin{equation}
\Psi_{L} (a) = C \Bigl(\frac{1}{1 - \frac{\Lambda}{3} a^2}
\Bigr)^{1/4} e^{\frac{1}{\hbar} S_g (a)},
\end{equation}
where $S_g$ is the gravitational instanton (\ref{grav ins}).
The wave function can be continued to the de Sitter region
\begin{equation}
\Psi_{L} (a) = C Bi \bigl( - \eta - \delta(\eta) \bigr),
\end{equation}
with the asymptotic form
\begin{equation}
\Psi_{L} (a) = \Bigl(\frac{1}{\frac{\Lambda}{3} a^2 - 1} \Bigr)^{1/4}
\cos \Bigl( \frac{1}{\hbar} S_g (a) + \frac{\pi}{4}\Bigr).
\end{equation}
Finally, Vilenkin's tunneling wave function is given by
\begin{equation}
\Psi_{V} (a) = C [Ai \bigl( \zeta - \delta(\zeta) \bigr) - i Bi
\bigl( \zeta -\delta (\zeta) \bigr)],
\end{equation}
with the asymptotic form
\begin{equation}
\Psi_{V} (a) = C \Bigl(\frac{1}{1 - \frac{\Lambda}{3} a^2} \Bigr)^{1/4}
\Bigl[ \frac{1}{2} e^{ - \frac{1}{\hbar} S_g (a)} + e^{\frac{1}{\hbar}
S_g (a)} \Bigr].
\end{equation}
In the de Sitter region it becomes a purely outgoing wave function
\begin{equation}
\Psi_{V} (a) = C \eta^{1/2} H^{(1)}_{1/3}\Bigl( \frac{2}{3} \eta^{3/2}\Bigr).
\end{equation}

A few comments are in order. First, in the tunneling region,
Hartle-Hawking's, Linde's, and Vilenkin's  wave functions have the
asymptotic forms, $\Psi_{HH} \approx e^{ - S_g (a)/\hbar},
\Psi_{L} \approx e^{+ S_g/\hbar}$, and $\Psi_{V} \approx e^{+
S_g/\hbar} + e^{- S_g/\hbar}/2$, respectively. As $S_g (a)$ is a
decreasing function of $a$, Hartle-Hawking's wave function
increases from a minimum value whereas Linde's and Vilenkin's wave
functions decrease from a maximum values at $a = 0$. When $S_g$ is
large, Linde's wave function and Vilenkin's wave function are
dominated approximately by the same asymptotic form
$e^{S_g/\hbar}$. These forms are important in deriving
semiclassical gravity in the next sections. Second, in the de
Sitter region Hartle-Hawking's wave function and Linde's wave
functions have equal amplitudes (probabilities) for outgoing
(expanding) and incoming (collapsing) components whereas
Vilenkin's wave function has a purely outgoing component.

\section{Semiclassical Gravity}

The idea of deriving semiclassical gravity from quantum gravity is
rooted on the correspondence in certain circumstances
between quantum theory and classical theory.
In fact, the Schr\"{o}dinger equation
\begin{equation}
\Bigl[- \frac{\hbar^2}{2m} \nabla^2 + V(x) \Bigr] \psi (x) = E \psi (x),
\end{equation}
with a WKB solution of the form
\begin{equation}
\psi (x) = e^{\frac{i}{\hbar} S(x)},
\end{equation}
may lead, in the asymptotic limit $(1/\hbar \rightarrow \infty)$, to the
Hamilton-Jacobi equation
\begin{equation}
\frac{1}{2m} (\nabla S)^2 + V(x) = E. \label{hj eq}
\end{equation}
With the identification $p = \nabla S$ and $H = E$, Eq. (\ref{hj eq})
becomes nothing but the Hamiltonian
\begin{equation}
H = \frac{1}{2m} p^2 + V(x).
\end{equation}

The same idea may be applied to quantum cosmology. The WDW
equation for gravity $h_a$ coupled with a matter field $\phi$
takes in general the form
\begin{equation}
\Bigl[ - \frac{\hbar^2}{2m_p^2} \nabla^2 + m_p^2 V(h_a)
+ \hat{H}_m ( \pi_{\phi},
\phi, h_a ) \Bigr] \Psi(h_a, \phi) = 0.
\end{equation}
Here $m_p^2 = 1/G$, and $\nabla^2 = G_{ab} \delta^2 / \delta h_a
\delta h_b$ is the Laplace-Beltrami operator on the superspace
$h_a$ of three geometries modulo diffeomorphisms, and $V(h_a)$ is
a potential from the three-curvature. Contrary to ordinary quantum
theory, there are two asymptotic parameters, $m_p \rightarrow
\infty$ and $1/\hbar \rightarrow \infty$. A simple analogy is to
write the wave functional as \cite{sem grav}
\begin{equation}
\Psi (h_a, \phi) = e^{ \frac{i}{\hbar} S_0(h_a)} \chi (\phi, h_a).
\end{equation}
In the limit of $1/\hbar \rightarrow \infty$, one obtains the
Einstein-Hamilton-Jacobi equation
\begin{equation}
\frac{1}{2m_p^2} (\nabla S_0)^2 + m_p^2 V(h_a) = 0, \label{ehj eq}
\end{equation}
and the time-dependent Schr\"{o}dinger equation
\begin{equation}
i \hbar \frac{\delta}{\delta t} \chi =
\hat{H}_m (\pi_{\phi}, \phi, h_a) \chi.
\end{equation}
Here a cosmological time is introduced along each classical
trajectory determined by the Einstein-Hamilton-Jacobi equation as
\begin{equation}
\frac{\delta}{\delta t} = \frac{1}{m_p^2} G_{a b}
\frac{\delta S_0}{\delta h_{(a}} \frac{\delta}{ \delta h_{b)}}. \label{cos time}
\end{equation}
However, this simple idea has a limitation that Eq. (\ref{ehj eq}) is a
pure gravity (vacuum solution) without the backreaction of matter.

To overcome this shortcoming, we may employ the Born-Oppenheimer
idea of separating gravity from matter, and the de Broglie-Bohm
interpretation of emergence of classical system with a quantum
potential \cite{kim-cqg}. That is, the wave function is written as
\begin{equation}
\Psi (h_a, \phi) = \psi (h_a) \Phi(\phi, h_a),
\end{equation}
where $\psi(h_a)$ is a function of the slow variable $h_a$ with
mass scale $m_p$ and $\Phi(\phi, h_a)$ is a function of the fast
variable $\phi$ with mass scale $m$. Then one may obtain  from the
WDW equation the gravity part equation
\begin{eqnarray}
\Bigl[ - \frac{\hbar^2}{2m_p^2} D^2 + m_p^2 V + \langle \hat{H}_m
\rangle - \frac{\hbar^2}{2m_p^2} \langle \bar{D}^2 \rangle \Bigr]
\psi = 0,
\end{eqnarray}
and the matter part equation
\begin{eqnarray}
- \frac{\hbar^2}{m_p^2} \frac{1}{\psi} (D \psi) \cdot (\bar{D}
\Phi) + ( \hat{H}_m - \langle \hat{H}_m \rangle ) \Phi \\ -
\frac{\hbar^2}{2m_p^2} ( \bar{D}^2 - \langle \bar{D}^2 \rangle)
\Phi = 0.
\end{eqnarray}
Here $D$ and $\bar{D}$ are covariant derivatives
\begin{equation}
D_a = \frac{\delta}{\delta h_a} + i A_a, \quad
\bar{D}_a = \frac{\delta}{\delta h_a} - i A_a,
\end{equation}
with an effective gauge potential
\begin{equation}
A_a (h_a) = - i \frac{\langle \Phi \vert \frac{\delta}{\delta h_a}
\vert \Phi \rangle}{\langle \Phi \vert \Phi \rangle},
\end{equation}
and $\langle \hat{H}_m \rangle$ and $\langle \bar{D}^2 \rangle$
are backreactions
\begin{eqnarray}
\langle \hat{H}_m \rangle = \frac{\langle \Phi \vert \hat{H}_m
\vert \Phi \rangle}{\langle \Phi \vert \Phi \rangle}, \quad
\langle \bar{D}^2 \rangle = \frac{\langle \Phi \vert \bar{D}^2
\vert \Phi \rangle}{\langle \Phi \vert \Phi \rangle}.
\end{eqnarray}
After removing the geometric phases
\begin{equation}
\psi = e^{- i \int A} \tilde{\psi}, \quad \Phi = e^{i \int A} \tilde{\Phi},
\end{equation}
and dropping tildes, and adopting the de Broglie-Bohm interpretation
\begin{equation}
\psi(h_a) = F(h_a) e^{\frac{i}{\hbar} S(h_a)},
\end{equation}
one finds the Einstein-Hamilton-Jacobi equation
\begin{equation}
\frac{1}{2m_p^2} (\nabla S)^2 + m_p^2 V + \langle \hat{H}_m \rangle = 0,
\label{ehj eq2}
\end{equation}
and the quantum field equation for the matter field
\begin{equation}
i \hbar \frac{\delta}{\delta t} \Phi = [\hat{H}_m - \langle \hat{H}_m
\rangle ] \Phi = 0.
\end{equation}
The cosmological time is defined with respect to the gravitational
action with the backreaction from the matter field
\begin{equation}
\frac{\delta}{\delta t} = \frac{1}{m_p^2} \nabla S \cdot \nabla.
\label{cos time2}
\end{equation}
In conclusion, the semiclassical gravity from quantum gravity
takes the form
\begin{eqnarray}
G_{\mu \nu} &=& \frac{8 \pi}{m_p^2} \Bigl[\langle \hat{T}_{\mu
\nu} \rangle
+ T_{\mu \nu}^Q \Bigr], \nonumber\\
i \hbar \frac{\delta}{\delta t} \vert \Phi \rangle &=&
\Bigl[ \hat{H}_m + \hat{H}_m^Q \Bigr] \vert \Phi \rangle  = 0,
\end{eqnarray}
where $ T_{\mu \nu}^Q$ and $\hat{H}_m^Q$ are quantum corrections.
For instance, the semiclassical gravity for the FRW cosmology is given by
\cite{kim-cqg}
\begin{eqnarray}
\Bigl(\frac{\dot{a}}{a} \Bigr)^2 + \Bigl(\frac{K}{a^2} -
\frac{\Lambda}{3} \Bigr) = \frac{8 \pi}{3 m_p^2 a^3} \Bigl[ H_{nn}
+ \frac{2 \pi \hbar^2}{3 m_p^2 a} \Bigl(U_{nn}^2 +
\frac{\dot{U}_{nn}}{\dot{a}}  \Bigr) \Bigr], \nonumber\\
\end{eqnarray}
where
\begin{eqnarray}
H_{nn} = \langle \Phi_n \vert \hat{H}_m \vert \Phi_n \rangle,\quad
U_{nn} = - \frac{1}{2} \frac{d(a \dot{a})/d\tau}{a \dot{a}^3}.
\end{eqnarray}

\section{Semiclassical Gravity for Tunneling Universe}

From now on we shall confine our attention to the FRW universe
with a minimal scalar field.
The action for a minimal massive scalar field
\begin{equation}
I_m = \int d^4 x (- g)^{1/2} \frac{1}{2} \Bigl[- g^{\mu \nu}
\partial_{\mu} \phi \partial_{\nu} \phi  - m^2 \phi^2 \Bigr],
\end{equation}
after the Fourier-decomposition, leads to the canonical Hamiltonian of
the form
\begin{eqnarray}
H_m = \sum_{\alpha} \frac{\pi_{\alpha}^2}{2 a^3} + \frac{a^3}{2}
\omega_{\alpha}^2 \phi_{\alpha}^2, \quad \omega_{\alpha} =
\Bigl(m^2 + \frac{\alpha^2}{a^2} \Bigr)^{1/2}. \label{ham}
\end{eqnarray}
Here $\alpha$ denotes the collective notation for the Fourier
or spherical harmonic modes
\begin{eqnarray}
\alpha^2 = \begin{cases} k^2, ~~~~~~~~~({\rm for~ a~ spatially~
flat~ FRW}), \cr j (j+2), ~~({\rm for~ a~ closed~ FRW}),
\end{cases}
\end{eqnarray}
and $\pi_{\alpha} = a^3  \dot{\phi}_{\alpha}$ is the canonical momentum. Thus the
Hamiltonian is equivalent to an infinite sum of time-dependent oscillators.
The Hamiltonian constraint for gravity plus matter system now becomes
\begin{equation}
H_g + H_m = 0,
\end{equation}
and leads to the WDW equation
\begin{equation}
\Bigl[- \frac{\hbar^2}{2m_p^2} \frac{\partial^2}{\partial a^2} +
V_g (a) - \frac{3a}{2} \hat{H}_m (\pi_{\alpha}, \phi_{\alpha}, a)
\Bigr] \Psi (a, \phi_{\alpha}) = 0. \label{wdw mat}
\end{equation}

Without complication of a  backreaction of $\phi$ and loss of
generality, we assume that only a few modes be present in Eq.
(\ref{wdw mat}). We then define the cosmological time (\ref{cos
time}) along the classical trajectory of an expanding universe
with the gravity part of wave function
\begin{equation}
\psi_{\rm exp} (a) = F(a) e^{\frac{i}{\hbar} S_g (a)}
\end{equation}
as
\begin{equation}
\frac{\partial}{\partial t} = \frac{2}{3 \pi m_p^2 a}
\frac{\partial S_g}{\partial a} \frac{\partial}{\partial a}.
\label{frw time1}
\end{equation}
Then the field equation becomes
\begin{equation}
i \hbar \frac{\partial}{\partial t} \Phi (\phi_{\alpha}, a(t)) =
\hat{H}_m (\phi_{\alpha}, a(t)) \Phi (\phi_{\alpha}, a(t)).
\label{frw eq1}
\end{equation}
This expanding wave function can be analytically continued along a
semicircle in the upper half of a complex plane to the gravity
wave function in a tunneling region
\begin{equation}
\psi_{HH} = F(a) e^{- \frac{1}{\hbar} S_g (a)}.
\end{equation}
Then the cosmological time in the tunneling universe becomes
\begin{equation}
\frac{\partial}{\partial \tau} = \frac{2}{3 \pi m_p^2 a}
\frac{\partial S_g}{\partial a}, \label{tfrw time2}
\end{equation}
implying an imaginary time $t = i \tau$. The field
equation now takes the form
\begin{equation}
\hbar \frac{\partial}{\partial \tau} \Phi (\phi_{\alpha}, a
(\tau)) = \hat{H}_m (\phi_{\alpha}, a(\tau)) \Phi (\phi_{\alpha},
a(\tau)). \label{tfd eq1}
\end{equation}
Note that the quantization rule in the tunneling universe
(Euclidean geometry) changes to
\begin{equation}
[\hat{\phi}_{\alpha}, \hat{\pi}_{\beta} ] = i \hbar \delta_{\alpha
\beta} \Longrightarrow [\hat{\phi}_{\alpha}, \hat{\pi}_{\beta} ] =
- \hbar \delta_{\alpha \beta}.
\end{equation}
In the coordinate representation, $\hat{\pi}_{\alpha} = \hbar
\partial/\partial \phi_{\alpha}$,
the field equation in the tunneling universe becomes
\begin{equation}
\hbar \frac{\partial}{\partial \tau} \Phi (\phi_{\alpha}, a
(\tau)) = \Big[- \frac{\hbar^2}{2 a^3} \frac{\partial^2}{\partial
\phi_{\alpha}^2} + \frac{a^3}{2} \omega_{\alpha}^2 \phi_{\alpha}^2
\Bigr] \Phi (\phi_{\alpha}, a(\tau)).
\end{equation}

In passing, we note that the time-reversal operation of field
equations can be naturally understood in quantum cosmology. The
wave function for the collapsing universe
\begin{equation}
\psi_{\rm col} (a) = F(a) e^{- \frac{i}{\hbar} S_g (a)},
\end{equation}
with respect to the cosmological time (\ref{frw time1}), leads to
another field equation
\begin{equation}
- i \hbar \frac{\partial}{\partial t} \tilde{\Phi} (\phi_{\alpha},
a(t)) = \hat{H}_m (\phi_{\alpha}, a(t)) \tilde{\Phi}
(\phi_{\alpha}, a(t)). \label{frw eq2}
\end{equation}
In fact, Eq. (\ref{frw eq2}) is the time-reversal, $\tilde{\Phi}
(t) = \Phi^* (- t)$, of the field equation (\ref{frw eq1}) from
the expanding universe. Similarly, in the tunneling universe the
gravity wave function
\begin{equation}
\psi = F(a) e^{+ \frac{1}{\hbar} S_g (a)},
\end{equation}
leads to another field equation
\begin{equation}
- \hbar \frac{\partial}{\partial \tau} \tilde{\Phi}
(\phi_{\alpha}, a (\tau)) = \hat{H}_m (\phi_{\alpha}, a(\tau))
\tilde{\Phi} (\phi_{\alpha}, a(\tau)). \label{tfd eq2}
\end{equation}
Note that the field Equation (\ref{tfd eq2}), being real, is also
the time-reversal of Eq. (\ref{tfd eq1}). However, this equation
is necessary to equip the quantum field theory with an inner
product
\begin{equation}
\langle \langle \Phi_2 (\phi_{\alpha}, \tau) \vert \hat{O} \vert
\Phi_1 (\phi_{\alpha}, \tau) \rangle = \int d\phi_{\alpha}
\tilde{\Phi}_2 (\phi_{\alpha}, \tau) \hat{O} \Phi_1
(\phi_{\alpha}, \tau). \label{t inner}
\end{equation}
It should be remarked that, with the time convention (\ref{tfrw
time2}), Hartle-Hawking's wave function leads to Eq. (\ref{tfd
eq1}) whereas Linde's wave function leads to Eq. (\ref{tfd eq2}).
These two equations complement each other with respect to the
inner product (\ref{t inner}).

\section{Quantum Field Theory: Lorentzian vs. Tunneling Universes}

In this second part we study and compare the semiclassical gravity
in a tunneling universe and a pure Lorentzian universe. We first
introduce a scheme in the Lorentzian universe to find the Fock
space of the minimal massive scalar field in Sec. III, and then
extend the scheme to the tunneling universe to obtain the
corresponding Fock space.

\subsection{Fock Space in Lorentzian Geometry}

The Hamiltonian (\ref{ham}) of the minimal scalar field consists
of time-dependent oscillators that depend on time through $a$. The
invariant method \cite{lewis} turns out a powerful tool in finding
various quantum states for such a time-dependent quadratic
Hamiltonian. Following Ref. \cite{kim-wave}, we introduce the
time-dependent annihilation operator
\begin{eqnarray}
\hat{A}_{\alpha} (t) &=& \frac{i}{\sqrt{\hbar}} [
\varphi_{\alpha}^*(t) \hat{\pi}_{\alpha} - a^3 (t)
\dot{\varphi}_{\alpha}^* (t) \hat{\phi}_{\alpha} ], \label{cr-an}
\end{eqnarray}
and the creation operator $\hat{A}^{\dagger}_{\alpha}(t)$, the
Hermitian conjugate of $\hat{A}_{\alpha} (t)$, and require them to
satisfy the quantum Liouville-von Neumann equation
\begin{eqnarray}
i \hbar \frac{\partial}{\partial t} \hat{A}_{\alpha} (t) + [
\hat{A}_{\alpha} (t), \hat{H}_{\alpha} (t) ] = 0. \label{lvn eq}
\end{eqnarray}
Here an overdot denoting a derivative with $t$.  Then the
auxiliary field $\varphi_{\alpha}$ satisfies the classical
equation for each mode,
\begin{equation}
\ddot{\varphi}_{\alpha} + 3\frac{\dot{a}}{a}
\dot{\varphi}_{\alpha} + \omega_{\alpha}^2 \varphi_{\alpha} = 0.
\label{cl eq}
\end{equation}
In fact, these operators satisfy the equal-time commutator
\begin{equation}
[ \hat{A}_{\alpha} (t), \hat{A}_{\beta}^{\dagger} (t) ] =
\delta_{\alpha \beta},
\end{equation}
when $\varphi_{\alpha}$ is restricted by the Wronskian condition
\begin{equation}
a^3 ( \varphi_{\alpha} \dot{\varphi}_{\alpha}^* -
\varphi_{\alpha}^* \dot{\varphi}_{\alpha} ) = i. \label{wr con}
\end{equation}

The number state of the number operator, $ \hat{N}_{\alpha} (t) =
\hat{A}_{\alpha}^{\dagger} (t) \hat{A}_{\alpha} (t)$,
\begin{eqnarray}
\hat{N}_{\alpha} (t) \vert n_{\alpha}, t \rangle = n_{\alpha}
\vert n_{\alpha}, t \rangle, \label{num st}
\end{eqnarray}
is an exact solution to Eq. (\ref{frw eq1}). Once we select a
solution $\varphi_{{\alpha}0}$ satisfying both Eqs. (\ref{cl eq})
and (\ref{wr con}), we find the gaussian and its excited states as
\cite{kim-wave}
\begin{eqnarray}
\Phi_{n_{{\alpha}0}} (\phi_{\alpha}, t) &=&
\Bigl(\frac{1}{2^{n_{\alpha}} (n_{\alpha})! \sqrt{2 \pi \hbar}
\rho_{{\alpha}0}} \Bigr)^{1/2} e^{- i \Theta_{{\alpha}0}
(n_{{\alpha}0} + 1/2)} \nonumber\\&& \times H_{n_{\alpha}}
\Bigl(\frac{\phi_{\alpha}}{\sqrt{2 \hbar} \rho_{{\alpha}0}} \Bigr)
\exp \Bigl[\frac{i a^3 \dot{\varphi}_{{\alpha}0}^*}{2 \hbar
\varphi_{{\alpha}0}^*} \phi_{\alpha}^2\Bigr], \label{wav}
\end{eqnarray}
where $\varphi_{{\alpha}0} = \rho_{{\alpha}0} e^{- i
\Theta_{{\alpha}0}}$. The most general gaussian state and its
excited states can be obtained from a more general complex
solution of the form
\begin{equation}
\varphi_{{\alpha} \nu} (t) = \mu_{\alpha} \varphi_{{\alpha}0} (t)
+ \nu_{\alpha} \varphi_{{\alpha}0}^* (t), \label{most sol}
\end{equation}
where $\mu_{\alpha}$ and $\nu_{\alpha}$ are time-independent
parameters defined as
\begin{eqnarray}
\mu_{\alpha} = \cosh r_{\alpha}, \quad \nu_{\alpha} = e^{- i
\theta_{\alpha}} \sinh r_{\alpha}. \label{sq par}
\end{eqnarray}
In fact, the time-dependent annihilation and creation operators in
Eq. (\ref{cr-an}) defined in terms of $\varphi_{{\alpha}0}$ and
$\varphi_{\alpha \nu}$ are related through the Bogoliubov
transformation
\begin{eqnarray}
\hat{A}_{{\alpha} \nu} (t) &=& (\cosh r_{\alpha})
\hat{A}_{{\alpha}0} (t) - (e^{ i \theta_{\alpha}} \sinh
r_{\alpha}) \hat{A}^{\dagger}_{{\alpha}0} (t),
\nonumber\\
\hat{A}^{\dagger}_{{\alpha} \nu} (t) &=& (\cosh r_{\alpha})
\hat{A}^{\dagger}_{{\alpha}0} (t) - (e^{- i \theta_{\alpha}} \sinh
r_{\alpha}) \hat{A}_{{\alpha}0} (t). \nonumber\\
\label{bog tran}
\end{eqnarray}
The $\nu$-dependent number state that is obtained by replacing
$\varphi_{\alpha 0}$ with $\varphi_{\alpha \nu}$ in Eq.
(\ref{wav}) is a squeezed state of the number state (\ref{wav})
\cite{kim04}.

\subsection{Fock Space in Euclidean Geometry}

In the Euclidean geometry of the tunneling universe, the
non-unitary evolution equation (\ref{tfd eq1}) has also a pair of
the invariant operators of the form \cite{kim-euc}
\begin{eqnarray}
\hat{A}_{{\alpha}+} (\tau) &=& - \frac{1}{\sqrt{\hbar}} [
\varphi_{{\alpha}+}(\tau) \hat{\pi}_{\alpha} - a^3 (\tau)
\dot{\varphi}_{{\alpha}+} (\tau)
\hat{\phi}_{\alpha}], \nonumber\\
\hat{A}_{{\alpha}-} (\tau) &=&  \frac{1}{\sqrt{\hbar}} [
\varphi_{{\alpha}-}(\tau) \hat{\pi}_{\alpha} - a^3 (\tau)
\dot{\varphi}_{{\alpha}-} (\tau) \hat{\phi}_{\alpha} ], \label{euc
cr-an}
\end{eqnarray}
where an overdot denotes now a derivative with respect to $\tau$
and $\varphi_{{\alpha} \pm}$ are two independent solutions of the
classical field equation
\begin{equation}
\ddot{\varphi}_{{\alpha} \pm} + 3\frac{\dot{a}}{a}
\dot{\varphi}_{{\alpha} \pm} - \omega_{\alpha}^2 \varphi_{{\alpha}
\pm} = 0. \label{euc cl eq}
\end{equation}
These operators satisfy the quantum Liouville-von Neumann equation
in the Euclidean geometry
\begin{equation}
\hbar \frac{\partial}{\partial \tau} \hat{A}_{{\alpha} \pm} (\tau)
+ [\hat{A}_{{\alpha} \pm} (\tau), \hat{H}_{\alpha} (\tau)] = 0.
\end{equation}
With the Wronskian condition
\begin{equation}
a^3 ( \varphi_{{\alpha}-} \dot{\varphi}_{{\alpha}+} -
\varphi_{{\alpha}+} \dot{\varphi}_{{\alpha}-}) = 1 \label{euc wr
con}
\end{equation}
imposed, the equal time commutation relations
\begin{equation}
[\hat{A}_{{\alpha} -} (\tau), \hat{A}_{{\beta}+}(\tau)] =
\delta_{\alpha \beta}
\end{equation}
are satisfied. Note then that $\hat{A}_{{\alpha}-}(\tau)$ is an
analytical continuation of the annihilation operator
$\hat{A}_{\alpha} (t)$ and $\hat{A}_{{\alpha}+}(\tau)$ is another
continuation of the creation operator
$\hat{A}^{\dagger}_{\alpha}(t)$ of the Lorentzian geometry.

The zero eigenvalue ket-state of $\hat{A}_{{\alpha} -} (\tau)$,
\begin{equation}
\hat{A}_{{\alpha} -} (\tau) \vert 0_{\alpha}, \tau \rangle = 0,
\end{equation}
leads to a gaussian wave function
\begin{equation}
\Phi_{{\alpha}0} (\phi_{\alpha}, \tau) = \Bigl(\frac{1}{\sqrt{2
\pi \hbar} \varphi_{{\alpha}-}} \Bigr)^{1/2} \exp \Bigl[\frac{a^3
\dot{\varphi}_{{\alpha}-}}{2 \hbar \varphi_{{\alpha}-}}
\phi_{\alpha}^2 \Bigr]. \label{euc vac wav}
\end{equation}
As $\hat{A}_{\alpha \pm} (\tau)$ map one exact solution into
another, we obtain an excited ket-state given by
\begin{equation}
\vert n_{\alpha}, \tau \rangle = \frac{1}{\sqrt{n_{\alpha}!}}
\bigl( \hat{A}_{{\alpha}+} (\tau) \bigr)^{n_{\alpha}} \vert
0_{\alpha}, \tau \rangle,
\end{equation}
whose wave function is \cite{kim-page04}
\begin{eqnarray}
 \Phi_{n_{\alpha}} (\phi_{\alpha}, \tau) =
\Bigl(\frac{1}{2^{n_{\alpha}} (n_{\alpha})! \sqrt{2 \pi \hbar
\varphi_{{\alpha}+} \varphi_{{\alpha}-}}} \Bigr)^{1/2}
~~~~~~~~~~~~~~~~~~ \nonumber\\
\times
 \Bigl( \frac{\varphi_{{\alpha}+}}{\varphi_{{\alpha}-}}\Bigr)^{
 (n_{\alpha} + 1/2)/2}  H_{n_{\alpha}} \Bigl( \frac{\phi_{\alpha}}{\sqrt{2
\hbar \varphi_{{\alpha}+} \varphi_{{\alpha}-}}} \Bigr) \exp
\Bigl[\frac{a^3 \dot{\varphi}_{{\alpha}-}}{2 \hbar
\varphi_{{\alpha}-}} \phi_{\alpha}^2 \Bigr]. \nonumber\\
\label{euc num wav}
\end{eqnarray}
One can also show through direct calculation that the wave
functions (\ref{euc num wav}) indeed satisfy  Eq. (\ref{tfd eq1}).

The bra-state, which is needed for the inner product of quantum
states in Eq. (\ref{t inner}), may be defined by the equation
\begin{equation}
\langle \langle 0_{\alpha}, \tau \vert
\hat{A}^{\dagger}_{{\alpha}+} (\tau) = 0,
\end{equation}
and
\begin{equation}
\langle \langle n_{\alpha}, \tau \vert =
\frac{1}{\sqrt{n_{\alpha}!}} \langle 0_{\alpha}, \tau \vert
\bigl(\hat{A}^{\dagger}_{{\alpha}-} (\tau) \bigr)^{n_{\alpha}}.
\end{equation}
Here the Hermitian conjugates of $\hat{A}^{\dagger}_{{\alpha} \pm}
(\tau)$ are obtained by substituting $- \hat{\pi}_{\alpha}$ for
$\hat{\pi}_{\alpha}$ in Eq. (\ref{euc cr-an}). In fact,
$\hat{A}^{\dagger}_{{\alpha} \pm} (\tau)$ satisfy the
Liouville-von Neumann equation for the field equation (\ref{tfd
eq2})
\begin{equation}
- \hbar \frac{\partial}{\partial \tau} \hat{A}^{\dagger}_{{\alpha}
\pm} (\tau) + [\hat{A}^{\dagger}_{{\alpha} \pm} (\tau),
\hat{H}_{\alpha} (\tau)] = 0.
\end{equation}
The wave function for the number bra-state now takes the form
\cite{kim-page04}
\begin{eqnarray}
\tilde{\Phi}_{n_{\alpha}} (\phi_{\alpha}, \tau) =
\Bigl(\frac{1}{2^{n_{\alpha}} (n_{\alpha})! \sqrt{2 \pi \hbar
\varphi_{{\alpha}+} \varphi_{{\alpha}-}}}
\Bigr)^{1/2}~~~~~~~~~~~~~~~~~~~
 \nonumber\\
\times \Bigl(
\frac{\varphi_{{\alpha}-}}{\varphi_{{\alpha}+}}\Bigr)^{(n_{\alpha}
+ 1/2)/2} H_{n_{\alpha}} \Bigl( \frac{\phi_{\alpha}}{\sqrt{2 \hbar
\varphi_{{\alpha}+} \varphi_{{\alpha}-}}} \Bigr) \exp \Bigl[-
\frac{a^3 \dot{\varphi}_{{\alpha}+}}{2 \hbar
\varphi_{{\alpha}+}} \phi_{\alpha}^2 \Bigr]. \nonumber\\
\end{eqnarray}
The inner product satisfies the orthonormality
\begin{equation}
\langle \langle m_{\alpha}, \tau \vert n_{\alpha}, \tau \rangle =
\int d \phi_{\alpha} \tilde{\Phi}_{m_{\alpha}} (\phi_{\alpha},
\tau) \Phi_{n_{\alpha}} (\phi_{\alpha}, \tau) = \delta_{m_{\alpha}
n_{\alpha}}.
\end{equation}

\subsection{Mean Number of Created Particles in Tunneling Universe}

As $\hat{A}_{\alpha \pm} (\tau)$ depend on $\tau$ through
$a(\tau)$, there is a dynamical squeezing of an initial state. It
is the pair $\hat{A}_{\alpha -} (\tau)$ and
$\hat{A}^{\dagger}_{\alpha +} (\tau)$ that satisfy the Bogoliubov
transformation between operators at two different times:
\begin{eqnarray}
\hat{A}_{{\alpha}-} (\tau) &=& \mu_{\alpha} (\tau, \tau_0)
\hat{A}_{{\alpha}-} (\tau_0) + \nu_{\alpha} (\tau, \tau_0)
\hat{A}_{{\alpha}+}^{\dagger} (\tau_0),
\nonumber\\
\hat{A}_{{\alpha} +}^{\dagger} (\tau) &=& \tilde{\mu}_{\alpha}
(\tau, \tau_0) \hat{A}^{\dagger}_{{\alpha}+} (\tau_0) +
\tilde{\nu}_{\alpha} (\tau, \tau_0) \hat{A}_{{\alpha} -} (\tau_0).
\nonumber\\ \label{euc bog}
\end{eqnarray}
Though $[\hat{A}_{{\alpha}-} (\tau), \hat{A}_{{\alpha}
+}^{\dagger} (\tau)] \neq 1$, we can show by direct calculation
that the Bogoliubov relation holds
\begin{equation}
\tilde{\mu}_{\alpha} \mu_{\alpha} - \tilde{\nu}_{\alpha}
\nu_{\alpha} = 1.
\end{equation}
The initial vacuum state at $\tau_0$ contains particles at $\tau$
of the amount
\begin{equation}
\langle \langle 0_{\alpha}, \tau_0 \vert
\hat{A}^{\dagger}_{{\alpha} +} (\tau) \hat{A}_{{\alpha}-} (\tau)
\vert 0_{\alpha}, \tau_0 \rangle = \tilde{\nu}_{\alpha} (\tau,
\tau_0) \nu_{\alpha} (\tau, \tau_0). \label{euc mean1}
\end{equation}
The vacuum state at $\tau$ can be expressed as a superposition of
the number states at $\tau_0$. The probability is now given by
projecting into number states at $\tau_0$ as
\begin{equation}
P_{n_{\alpha}} = \langle \langle 0_{{\alpha}}, \tau \vert
n_{\alpha}, \tau_0 \rangle \langle \langle n_{\alpha}, \tau_0
\vert 0_{{\alpha}}, \tau \rangle.
\end{equation}
It should be remarked that $\langle \langle 0_{{\alpha}}, \tau
\vert n_{\alpha}, \tau_0 \rangle$ {\it is not} a mere complex
conjugate of $\langle \langle n_{\alpha}, \tau_0 \vert 0_{\alpha},
\tau \rangle$, which is true in the Lorentzian case. The
nonvanishing values are given by
\begin{eqnarray}
\langle \langle 2 n_{\alpha}, \tau_0 \vert 0_{\alpha}, \tau
\rangle &=& \frac{1}{\mu_{\alpha}^{1/2}}
\Bigl(\frac{(2n_{\alpha})!}{(n_{\alpha}!)^2} \Bigr)^{1/2}
\Bigl(- \frac{\nu_{\alpha}}{2 \mu_{\alpha}} \Bigr)^{n_{\alpha}}, \label{ket}\\
\langle \langle 0_{\alpha}, \tau \vert 2 n_{\alpha}, \tau_0
\rangle &=& \frac{1}{\tilde{\mu}_{\alpha}^{1/2}}
\Bigl(\frac{(2n_{\alpha})!}{(n_{\alpha}!)^2} \Bigr)^{1/2} \Bigl(-
\frac{\tilde{\nu}_{\alpha}}{2 \tilde{\mu}_{\alpha}}
\Bigr)^{n_{\alpha}}.\label{bra}
\end{eqnarray}
Using \cite{cal}, we find the mean number of created particles
\begin{equation}
\bar{n}_{\alpha} = \sum_{n_{\alpha}} 2n_{\alpha}
\Bigl[\frac{1}{(\tilde{\mu}_{\alpha}\mu_{\alpha})^{1/2}}
\frac{(2n_{\alpha})!}{(n_{\alpha}!)^2}
\Biggl(\frac{\tilde{\nu}_{\alpha} \nu_{\alpha}}{4
\tilde{\mu}_{\alpha} \mu_{\alpha}} \Biggr)^{n_{\alpha}} \Bigr] =
\tilde{\nu}_{\alpha} \nu_{\alpha}. \label{euc mean2}
\end{equation}
These two results are consistent with each other. The amplitude
squared of Eq. (\ref{ket}) gives a result different from (\ref{euc
mean1}) and (\ref{euc mean2}). Thus a complete quantum theory in
the Euclidean geometry requires both the ket- and bra-states,
which are not complex conjugates of each other.

The wave function should be continuous across the boundary between
the Lorentzian and Euclidean  geometries, $\Phi_{\alpha} (t_0) =
\Phi_{\alpha} (\tau_0)$. The continuity of the auxiliary fields,
$\varphi_{\alpha} (t_0) = \varphi_{\alpha} (\tau_0)$, implies the
continuity of wave function. First, we select the adiabatic
solution to Eq. (\ref{euc cl eq}) of the form
\begin{equation}
\varphi_{{\alpha} \pm} (\tau) \approx \frac{D_{{\alpha} \pm}
(\tau_0)}{ \sqrt{2 a^3 (\tau) \omega_{\alpha} (\tau)}} e^{ \pm
\int^{\tau}_{\tau_0} \omega_{\alpha}}, \label{euc ad sol}
\end{equation}
where
\begin{eqnarray}
D_{{\alpha} \pm} (\tau_0) = \sqrt{2 a^3 (\tau_0) \omega_{\alpha}
(\tau_0)} \varphi_{{\alpha} \pm} (\tau_0).
\end{eqnarray}
Then the wave function for the adiabatic vacuum state is
\begin{eqnarray}
\Phi_{{\alpha}0} (\phi_{\alpha}, \tau) \approx \Bigl(\frac{a^3
\omega_{\alpha}}{\pi \hbar} \Bigr)^{1/4} \exp \Bigl[- \frac{a^3
\omega_{\alpha}}{2 \hbar} \phi_{\alpha}^2 \Bigr]. \label{euc ad
wav}
\end{eqnarray}
Note that Eq. (\ref{euc ad wav}) can be obtained through an
analytical continuation of the adiabatic vacuum state in the
Lorentzian geometry \cite{ad vac}. Now, the mean number of created
particles can be calculated by expanding the wave function with
respect to the number states at the moment of entering the
tunneling universe
\begin{equation}
\Phi_{{\alpha}0} (\phi_{\alpha}, \tau) = \sum_{n_{\alpha}}
c_{n_{\alpha}} \Phi_{n_{\alpha}} (\phi_{\alpha}, \tau_0) =
\sum_{n_{\alpha}} c_{n_{\alpha}} \Phi_{n_{\alpha}} (\phi_{\alpha},
t_0).
\end{equation}
We thus see that the adiabatic vacuum state minimizes particle
creation agreeing with the result by Vilenkin {\it et al}
\cite{vilenkin}.

Second, there is one complex-parameter squeezed gaussian state
which can be obtained from a superposition of adiabatic solutions
\begin{eqnarray}
\varphi_{{\alpha} \nu \pm} (\tau) \approx \frac{1}{\sqrt{2 a^3
\omega_{\alpha}}} \Bigl(\cosh r_{\alpha} e^{ \pm
\int^{\tau}_{\tau_0} \omega_{\alpha}} D_{{\alpha}\pm} (\tau_0) \nonumber\\
+ e^{ \mp i \theta_{\alpha}} \sinh r_{\alpha} e^{\mp
\int^{\tau}_{\tau_0} \omega_{\alpha}} D_{{\alpha}\mp} (\tau_0)
\Bigr).  \label{euc gen ad sol}
\end{eqnarray}
This leads to the squeezed gaussian wave function
\begin{eqnarray}
\Phi_{{\alpha}0} (\phi_{\alpha}, \tau) = {\cal N}_{{\alpha}+} \exp
\Biggl[ \frac{ a^3 \omega_{\alpha}}{2 \hbar} B_{{\alpha} \nu}
\phi_{\alpha}^2\Biggr], \label{euc sq ad wav}
\end{eqnarray}
where ${\cal N}_{{\alpha}+}$ is a normalization factor and
\begin{eqnarray}
B_{{\alpha} \nu} = \frac{ e^{ - i \theta_{\alpha} } \sinh
r_{\alpha} D_{{\alpha}-} - \cosh r_{\alpha} e^{ - 2
\int^{\tau}_{\tau_0} \omega_{\alpha}} D_{{\alpha}+}}{ e^{ - i
\theta_{\alpha}} \sinh r_{\alpha} D_{{\alpha}-} + \cosh r_{\alpha}
e^{ - 2 \int^{\tau}_{\tau_0} \omega_{\alpha}} D_{{\alpha}+}}.
\nonumber\\
\end{eqnarray}
Note that
\begin{equation}
{\rm Re} (B_{{\alpha}\nu}) > 0  \Longleftrightarrow  \Bigl|
\frac{D_{{\alpha}-}}{D_{{\alpha}+}} \Bigr| \tanh r_{\alpha} > e^{
- 2 \int^{\tau}_{\tau_0} \omega_{\alpha} }. \label{cat con}
\end{equation}
Thus for a sufficiently long duration of tunneling, even a small
squeezing parameter $r_{\alpha}$ eventually leads to an
exponentially growing state at the end of tunneling. Therefore,
the squeezed state satisfying the inequality (\ref{cat con}) forms
a one-parameter family of gaussian wave function that leads to
catastrophic particle production.

\section{Conclusion}

Using a quantum cosmological model for the
Friedmann-Robertson-Walker universe with a cosmological constant,
we have prescribed three leading proposals of Hartle-Hawking's
no-boundary wave fucntion, Linde's wave function, and Vilenkin's
tunneling wave function. We also derived in a general context the
semiclassical gravity from quantum gravity based on the
Wheeler-DeWitt equation. Further, the difference of semiclassical
gravity was discussed for the three leading wave functions.

Particle creation has been studied for a minimal massive scalar
field in a tunneling universe from the semiclassical gravity point
of view. Using the invariant method we found the Fock space of the
scalar field in the Lorentzian and Euclidean geometries of the
tunneling universe. It was found that the adiabatic vacuum state
minimizes particle creation in the tunneling universe whereas its
one-parameter squeezed gaussian states lead to catastrophic
particle creation for a sufficiently long duration of tunneling
regime. Thus the debate of catastrophic particle creation reduces
to the selection of gaussian states.

It would be interesting to compare the result of this paper with
another cosmological model where a tunneling region is sandwiched
between two asymptotically static Lorentzian regions. In this new
model the initial vacuum state evolves through the tunneling
region into a squeezed gaussian state always with catastrophic
particle creation whereas the evolution without the tunneling
region suppresses particle creation \cite{kim-page04}.

\acknowledgements

The author would like to thank Don N. Page for many useful
discussions and, in particular, for the collaboration on particle
creation in tunneling universes. The work was supported by the
Korea Astronomy Observatory.

\end{document}